\documentclass[12pt,preprint]{aastex}

\usepackage{emulateapj5}

\shorttitle{EFFECT OF DIFFERENTIAL ROTATION}
\shortauthors{MORRISON, BAUMGARTE \& SHAPIRO}

\begin{document}
\title{Effect of Differential Rotation on the Maximum Mass of 
Neutron Stars: Realistic Nuclear Equations of State}
\author{Ian A.~Morrison\altaffilmark{1},
Thomas W.~Baumgarte \altaffilmark{1,2} and
Stuart L.~Shapiro \altaffilmark{2,3}}
%
\affil
{\altaffilmark{1} 
Department of Physics and Astronomy, Bowdoin College,
Brunswick, ME 04011}
\affil
{\altaffilmark{2} 
Department of Physics, University of Illinois at Urbana-Champaign,
Urbana, IL 61801}
\affil
{\altaffilmark{3} 
Department of Astronomy and NCSA, University of Illinois at
Urbana-Champaign, Urbana, IL 61801}
%
\begin{abstract}
The merger of binary neutron stars is likely to lead to differentially
rotating remnants.  In this paper, we survey several cold nuclear
equations of state (EOSs) and numerically construct models of
differentially rotating neutron stars in general relativity.  For each
EOS we tabulate maximum allowed masses as a function of the degree of
differential rotation.  We also determine effective polytropic
indices, and compare the maximum allowed masses with those for the
corresponding polytropes.  We consistently find larger mass increases
for the polytropes, but even for the nuclear EOSs we typically find
maximum masses 50\% higher than the corresponding values for
nonrotating (TOV) stars.  We evaluate our findings for the six
observed binary neutron star (pulsar) systems, including the recently
discovered binary pulsar J0737-3039.  For each EOS we determine
whether their merger will automatically lead to prompt collapse to a
black hole, or whether the remnant can be supported against collapse
by uniform rotation (possibly as a supramassive star) or differential
rotation (possibly as a hypermassive star).  For hypermassive stars,
delayed collapse to a black hole is likely.  For the most recent EOSs
we survey the merger remnants can all be supported by rotation against
prompt collapse, but their actual fate will depend on the
nonequilibrium dynamics of the coalescence event.  Gravitational wave
observations of coalescing binary neutron stars may be able to
distinguish these outcomes -- no, delayed or prompt collapse -- and
thereby constrain possible EOSs.
\end{abstract}


%
%

\section{Introduction}

The remnant formed in the coalescence of binary neutron stars is
likely to be differentially rotating (see, e.g., the dynamical
simulations of Rasio \& Shapiro 1992, 1994, 1999; Shibata \& Ury\=u
2000, 2002; Faber, Rasio \& Manor 2001; Oechlin, Rosswog \&
Thielemann, 2002; Shibata, Taniguchi \& Ury\=u 2003; Faber,
Grandcl\'ement \& Rasio 2003; see also the review of Baumgarte and
Shapiro 2003).  It is likely that differential rotation will play an
important role in the dynamical stability of these remnants, since it
can be very effective in increasing the their maximum allowed mass
(Baumgarte, Shapiro \& Shibata 2000, hereafter BSS; Lyford, Baumgarte
\& Shapiro 2003, hereafter LBS).

Most neutron stars in binaries have individual gravitational masses
close to 1.4 $M_{\odot}$ (compare Table \ref{Table2} below).
Furthermore, most recent realistic nuclear equations of state predict
a maximum allowed mass for nonrotating neutron stars in the range of
about 1.7 -- 2.3 $M_{\odot}$ (compare Table \ref{Table1} below).
Taken together, these facts seem to suggest that the coalescence of
binary neutron stars would lead to prompt collapse to a black hole.
However, rotation, and especially differential rotation, can increase
the maximum allowed mass significantly\footnote{For coalescence from
the innermost stable circular orbit, shear and shock-induced thermal
pressure, which can also increase the maximum allowed mass, is found
to have a smaller effect.}.

The maximum allowed mass of {\em uniformly} rotating stars is limited
by the spin rate at which the fluid at the equator moves on a geodesic
(the Kepler limit); any further speed-up would lead to mass shedding.
Uniform rotation can therefore increase the maximum allowed mass by
about 20 \% at most for very stiff equations of state (Cook, Shapiro
\& Teukolsky, 1992, 1994a, 1994b, hereafter CST1, CST2 and CST3
respectively; see also Table \ref{Table1} below), which is not
sufficient to stabilize remnants of binary neutron star coalescence.
Uniformly rotating neutron stars with rest masses exceeding the
maximum allowed rest mass for non-rotating stars (for the same
equation of state) are referred to as {\em supramassive} neutron
stars.

{\em Differential} rotation, however, is much more effective in
increasing the maximum allowed mass. Unlike a uniformly rotating star,
the rotation rate at the core of a differentially rotating star is not
restricted to the maximum rotation rate at the equator, so that the
core can be supported by rapid rotation without the equator having to
exceed the Kepler limit.  This effect was demonstrated in Newtonian
gravitation by Ostriker, Bodenheimer \& Lynden-Bell (1966) for white
dwarfs, and in general relativity by BSS for $n=1$ polytropes.  BSS
also showed by way of illustration that stars with about 60\% more
mass than the maximum allowed mass of the corresonding nonrotating
star can be dynamically stable against both radial and nonaxisymmetric
modes.  BSS refer to differentially rotating equilibrium
configurations with rest masses exceeding the maximum rest mass of a
uniformly rotating star as {\em hypermassive} neutron stars.  LBS
generalized the equilibrium results of BSS to other polytropic indices
in the range $0.5 \leq n \leq 2.9$, and found that the largest
relative increases in the maximum allowed mass can be found for
polytropic indices close to $n=1$.  It is therefore not surprising
that the hypermassive binary neutron star remnants formed in the $n=1$
simulations of Shibata, Taniguchi \& Ury\=u (2003) do not collpase
promptly to black holes unless they exceed the maximum nonrotating
mass by more than about 70\%.  The merged remnants will be dynamically
stable on a dynamical timescale.  Viscous damping and magnetic braking
of differential rotation will likely occur on a secular timescale
(which is much greater than the dynamical timescale), leading to a
delayed collapse and a delayed burst of gravitational radiation (BSS;
Shapiro 2000; Cook, Shapiro \& Stevens 2003; Liu \& Shapiro 2003).

In this paper, we generalize the results of BSS and LBS to realistic
equations of state.  We construct differentially rotating neutron
stars for a sample of six cold nuclear equations of state, and
tabulate their maximum allowed masses as a function of the degree of
differential rotation.  In addition, we identify an ``effective''
polytropic index for each of these equations of state, and construct
differentially rotating polytropes for these indices.  We consistently
find larger mass increases for the polytropes than for the nuclear
equations of state, which is due to a drop in the maximum density for
a rotating star with a large degree of differential rotation.
However, even for nuclear equations of state we find increases in the
maximum allowed rest mass on the order of about 50\%, so that binary
neutron star coalescence should often result in hypermassive neutron
stars.  We explicitly predict the fate of the six double neutron star
binaries that have been identified to date, using the results of our
equilibrium model calculations.

The paper is organized as follows.  In Section \ref{numerics} we
briefly summarize the numerical method and discuss the nuclear
equations of state adopted in our calculations.  In Section
\ref{results} we present numerical results and compare our findings
for nuclear equations of state with those for polytropes.  We
summarize our findings and discuss consequences for the six observed
binary neutron star systems in Section \ref{summary}.  In Appendix
\ref{tables} we tabulate our numerical results.  We adopt
gravitational units and set $G = c = 1$.

%
%

\section{Constructing Numerical Models}
\label{numerics}

\subsection{Equilibrium Models of Differentially Rotating Stars}
\label{em}

As in BSS and LBS, we use a modified version of the numerical code of
CST to construct equilibrium models of differentially rotating neutron
stars in general relativity.  The code is based on similar numerical
methods developed by Hachisu (1986) and Komatsu, Eriguchi, \& Hachisu
(1989), and we refer to CST1 for details.

Constructing differentially rotating neutron star models requires
choosing a rotation law $F(\Omega) = u^t u_{\phi}$, where $u^t$ and
$u_{\phi}$ are components of the four-velocity $u^{\alpha}$ and
$\Omega$ is the angular velocity.  We follow CST1 and assume a
rotation law $F(\Omega) = A^2 (\Omega_c - \Omega)$, where the
parameter $A$ has units of length and where $\Omega_c$ is the central
value of the angular velocity.  Expressing $u^t$ and $u_{\phi}$ in
terms of $\Omega$ and metric potentials yields equation (42) of CST1,
or
\begin{equation}
\Omega = \frac{\Omega_c}{1 + \hat A^{-2} \hat r^2 \sin^2 \theta}
\end{equation}
in the Newtonian limit.  Here we have rescaled $A$ and $r$ in terms of
the equatorial radius $R_e$: $\hat A \equiv A/R_e$ and $\hat r \equiv
r/R_e$.  The parameter $\hat A$ is a measure of the degree of
differential rotation and determines the length scale over which
$\Omega$ changes.  Since uniform rotation is recovered in the limit
$\hat A^{-1} \rightarrow 0$, it is convenient to parametrize sequences
by $\hat A^{-1}$.  In the Newtonian limit the ratio between the
central and equatorial angular velocities $\Omega_c/\Omega_e$ is
related to $\hat A$ by $\Omega_c/\Omega_e = 1 + \hat A^{-2}$, but
for relativistic configurations this relation holds only
approximately.

We adopt this particular rotation law for convenience and for easy
comparison with many other authors who have assumed the same law.
However, as we pointed out in LBS, this rotation law approximates
reasonably well the angular velocity profile of binary neutron star
remnants in the fully relativistic dynamical simulations of Shibata \&
Uryu (2000, 2002) and the post-Newtonian simulations of Faber, Rasio,
\& Manor (2001; see also Faber, Grandcl\'ement \& Rasio 2003).  This
suggests that the above rotation law may provide a reasonable
parametrization of differential rotation profiles that one might
expect to find in binary neutron star merger remnants.

We modified the numerical algorithm of CST1 by fixing the maximum
interior density instead of the central density for each model.  This
change allows us to construct higher mass models in some cases, since
the central density does not always coincide with the maximum density
and hence may not specify a model uniquely.

For each equation of state and a given value of $\hat A$ we construct
a sequence of models for each value of the maximum density by starting
with a static, spherically symmetric star and then decreasing the
ratio of the polar to equatorial radius, ${\cal R} \equiv R_p/R_e$, in
decrements of 0.015.  This sequence ends when we reach mass shedding
(for large values of $\hat A$) or when the code fails to converge
(indicating the termination of equilibrium solutions) or when ${\cal R}
= 0$ (beyond which the star would become a toroid).  For each one of
these sequences, the maximum achieved mass is recorded.  We repeat
this procedure for different values of the maximum density, 
which yields the maximum achieved mass as a function of maximum
density.  The maximum of this curve is the maximum allowed mass for
this particular equation of state and the chosen value of $\hat A$.
In the Appendix we tabulate our numerical results for the equations of
state described in Section \ref{eos}, and for $\hat A^{-1} = 0, 0.3,
0.5, 0.7$ and 1.0.  Clearly, our maximum allowed masses are {\em lower
limits} in the sense that even higher mass models may exist for other
values of $\hat A$ or different differential rotation laws.

\subsection{Equations of State}
\label{eos}


\begin{table*}[b]
\caption{Maximum Mass Configurations}
\centerline{
\begin{tabular}{cccccccccc}
        \hline \hline
EOS &
$M^{\rm TOV}$ \tablenotemark{a} &
$M_0^{\rm TOV}$ &
$M^{\rm UNI}$ &
$M_0^{\rm UNI}$ &
$(\delta M_0 / M_0)^{\rm UNI}$ &
$M^{\rm DIF}$ &
$M_0^{\rm DIF}$ &
$(\delta M_0 / M_0)^{\rm DIF}$ &
$\hat A^{-1}$ \\
\hline
A \tablenotemark{b}    & 1.66 & 1.92 & 1.95 & 2.24 & 0.17 & 2.62 & 2.90 & 0.51 & 0.7 \\
D \tablenotemark{c}    & 1.65 & 1.89 & 1.95 & 2.22 & 0.18 & 2.71 & 3.02 & 0.60 & 0.7 \\
L \tablenotemark{d}    & 2.70 & 3.23 & 3.27 & 3.87 & 0.20 & 4.45 & 5.04 & 0.56 & 0.7 \\
UT \tablenotemark{e}   & 1.84 & 2.17 & 2.19 & 2.55 & 0.18 & 2.89 & 3.25 & 0.50 & 0.7 \\
FPS \tablenotemark{f}  & 1.80 & 2.10 & 2.12 & 2.45 & 0.17 & 2.69 & 3.08 & 0.46 & 0.5 \\
APR \tablenotemark{g} & 2.20 & 2.67 & 2.46 & 3.10 & 0.16 & 2.95 & 3.50 & 0.31 & 0.3 \\  
\hline
\end{tabular}}
\tablenotetext{a}{All masses are in units of Solar mass $M_{\odot}$.}
\tablenotetext{b}{Reid soft core; Pandharipande (1971).}
\tablenotetext{c}{Model V; Bethe and Johnson (1974).}
\tablenotetext{d}{Mean field; Pandharipande and Smith (1975).}
\tablenotetext{e}{UV14 $+$ TNI; Wiringa, Fiks, and Fabrocini (1988).}
\tablenotetext{f}{UV14 $+$ TNI; Lorenz, Ravenhall, and Pethick (1993).}
\tablenotetext{g}{A18 +$\delta v$ + UIX$^{*}$; Akmal, Pandharipande, and Ravenhall (1998).}
\label{Table1}
\end{table*}


\begin{figure*}
\plotone{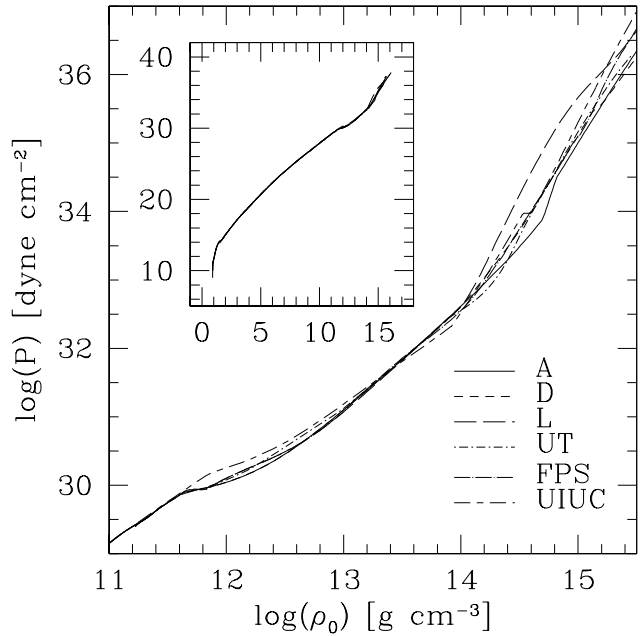} \figcaption[Fig1.eps]{Pressure as a function of
rest-mass density for the six nuclear equations of state adopted in
this paper (see text and Table \ref{Table1} for details).
\label{Fig1} }
\end{figure*}

In this paper we adopt six cold, nuclear equations of state (EOSs)
which are listed, together with numerical results, in Table
\ref{Table1}.  The first five, A, D, L, UT and FPS are adopted from
CST3 (who in turn adopted the labeling of Friedman, Ipser, \& Parker
(1986) for the first three EOSs).  The last EOS, APR, was adapted from
Akmal, Pandharipande \& Ravenhall (1998) by Ravenhall and
Pandharipande (with the smooth matching between low and high density
provided by G.~Cook).  In Figure \ref{Fig1} we plot the pressure $P$
as a function of the rest-mass density $\rho_0$ for these EOSs.

EOS A (Pandharipande 1971) models the interaction of neutrons at high
densities with a Reid soft-core potential.  EOS D is model V of Bethe
\& Johnson (1974).  In EOS L the nucleon interaction is modeled in
terms of a mean scalar field (Pandharipande, \& Smith 1975).  Both
EOSs UT (Wiringa, Fiks, \& Fabrocini 1988) and FPS (Lorenz, Ravenhall,
\& Pethick, 1993) are modern versions of an earlier EOS proposed by
Friedman \& Pandharipande (1981), which employs both two-body (U14) and
three-body nucleon interactions (TNI).  EOS UT improves the treatment
of matter at high densities, while FPS describes the interactions in
terms of a Skyrme model.  EOS APR (Akmal, Pandharipande, \& Ravenhall
1998) adopts a modern two-nucleon interaction (A18) together with
boost corrections, as well as the UIX three-nucleon potential.

The different descriptions of nucleon interactions affect the EOS only
at high densities.  At low densities ($\rho_0 \lesssim 10^4 \mbox{g
cm}^{-3}$), the EOSs employ the Feynman, Metropolis, \& Teller (1949)
EOS, joining onto the Baym, Pethick, \& Sutherland (1971) EOS up to
neutron drip at $\rho_0 \approx 4 \times 10^{11} \mbox{g cm}^{-3}$.
Above neutron drip, EOSs A, D join onto the Baym, Bethe, \& Pethick
(1971) EOS, while EOSs L and UT join onto the Negele \& Vautherin
(1973) EOS.  EOS APR joins onto the FPS EOS below a number density of
0.1 fm$^{-3}$ ($\rho_0 = 1.26 \times 10^{14} \mbox{g cm}^{-3}$),

All EOSs are read into our numerical code in tabular form, listing
rest-density $\rho_0$, the total energy density $\epsilon$, and the
pressure $P$ at discreet points.  Intermediate values of these
quantities are computed by interpolation.  Further details on the
numerical implementation of these EOSs, including the fitting and
tabulation, can be found in CST3.

%
%

\section{Results}
\label{results}

Our numerical results are summarized in Table \ref{Table1}, where we
list, for each equation of state, the nonrotating
(Tolman-Oppenheimer-Volkoff, TOV) maximum masses (both gravitational
mass $M^{\rm TOV}$ and rest mass $M_0^{\rm TOV}$), the maximum masses
for uniform rotation, and the maximum masses that we found in our
survey of differential rotation.  For convenience we also list
fractional mass increases with respect to the TOV maximum mass, as
well as the parameter $\hat A^{-1}$ for which the maximum mass was
encountered.  More detailed results, with results for all values of
$\hat A^{-1}$ considered in this paper, are tabulated in Appendix
\ref{tables}.

The fractional mass increases are typically in the order of 50\%.
These increases are still large enough to support the remnant of
a coalescing binary neutron star system in many cases (compare Section
\ref{summary}).  However, they are noticeably below the increases for
polytropic EOSs.  LBS found that the fractional mass increases depend
strongly on the polytropic index $n$, i.e.~the stiffness of the EOS.
For values of $n$ between 0.75 and 1.25, corresponding to moderately
stiff EOSs, they found relative mass increases exceeding 100\%.  It is
therefore somewhat surprising that the nuclear EOSs considered in this
paper yield significantly smaller numbers.


\begin{figure*}
\plotone{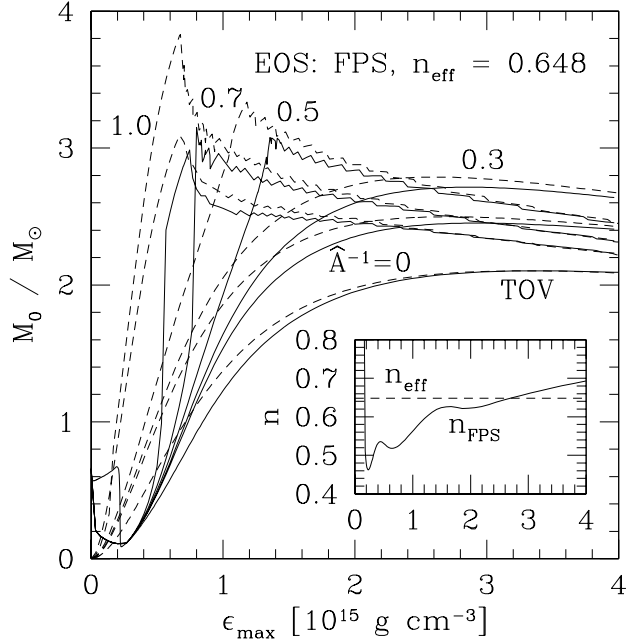} \figcaption[Fig2.eps]{The maximum rest mass as a
function of maximum density $\epsilon_{\rm max}$, for different values
of $\hat A$.  We include results for the FPS equation of state (solid
lines), as well as for an $n = 0.648$ polytrope (dashed lines).  The
agreement is quite good for large values of $\epsilon_{\rm max}$, but
less so for smaller values of $\epsilon_{\rm max}$ (for rotating as
well as non-rotating stars).  This can be understood in terms of the
polytropic index $n_{\rm FPS}$ (as computed from (\ref{pol_nucl});
solid line in the small panel) which drops significantly below $n_{\rm
eff}$ (as computed from central condensation; dashed line).  The
raggedness of some of the lines is a consequence of the finite
parameter step size in our sequences, and is a measure of the
numerical error in our results.\label{Fig2} }
\end{figure*}

To understand this result we construct models of differentially
rotating neutron stars for polytropic EOSs
\begin{equation} \label{poly_eos}
P = K \rho_0^{1+1/n},
\end{equation}
where $K$ a polytropic constant and $n$ the polytropic index.  For
each nuclear EOS we identify an ``effective'' polytropic index $n_{\rm
eff}$ as follows.  We first construct, for a nuclear EOS, the
nonrotating (TOV) maximum-mass model, and compute for this model the
central concentration $(\epsilon_c/\bar \epsilon)_{\rm nucl}$.  Here
$\epsilon_c$ is the central value of the energy density $\epsilon$,
and we define the average energy density from
\begin{equation}
\bar \epsilon \equiv \frac{3M}{4 \pi R^3},
\end{equation}
where $M$ is the gravitational mass and $R$ the circumferential
radius.  In Newtonian gravitation, this central condensation would
correspond to a unique value of $n$.  In general relativity, however,
this is no longer the case.  To identify the effective polytropic
index $n_{\rm eff}$, we therefore construct TOV maximum mass
configurations for a large sample of polytropic indices $n$, and
compute their central condensation $(\epsilon_c/\bar \epsilon)_{\rm
poly}$ as described above (compare Table 1 in LBS).  We then
interpolate between the polytropic values of the central condensation
$(\epsilon_c/\bar \epsilon)_{\rm poly}$ to the desired nuclear value
$(\epsilon_c/\bar \epsilon)_{\rm nucl}$, and thereby identify the
effective polytropic index $n_{\rm eff}$.  We fix the polytropic
constant $K$ in (\ref{poly_eos}) by setting the mass of the polytropic
maximum mass TOV model, which scales with $K^{n/2}$, equal to its
counterpart for the corresponding nuclear EOS.

We consistently find that the agreement between the nuclear EOS models
and the corresponding polytropes is quite good (to within 10\% or so)
for those models that have a maximum density $\epsilon_{\rm max}$
similar to the maximum mass TOV star.  However, for larger degrees of
differential rotation the maximum mass models shift to lower values of
$\epsilon_{\rm max}$ as the star become increasingly ``toroidal'' and
bloated perpendicular to the rotation axis.  We show this effect for
EOS FPS in Figure \ref{Fig2}, and we find qualitatively indentical
results for all other EOSs (see also Figure 1 in BSS).

This behavior can be understood by computing, as an alternative to our
construction above, a ``nuclear'' polytropic index from the slope of
the graphs in Fig.~\ref{Fig1}.  Using the adiabatic index
\begin{equation}
\Gamma = \frac{\partial \ln P}{\partial \ln \rho_0}
\end{equation}
we difine a nuclear polytropic index by
\begin{equation} \label{pol_nucl}
n_{\rm nucl} = \frac{1}{\Gamma - 1}. 
\end{equation}
Clearly this index is a function of density, and its value reflects
different interactions dominating different density regimes.  Across
phase transitions, $n_{\rm nucl}$ can even change discontinously.  We
include a plot of $n_{\rm FPS}$ as a function of the density
$\epsilon$ in the inserted panel in Figure \ref{Fig2}.


\begin{figure*}
\plotone{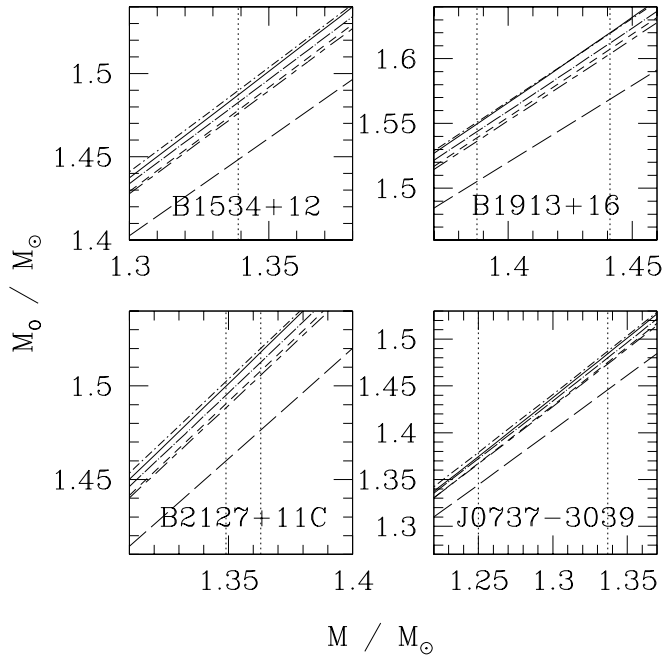} \figcaption[Fig3.eps]{Rest mass $M_0$ as a function
of gravitational mass $M$ for the six nuclear equations of state
(labeled as in Fig.~\ref{Fig1}).  The vertical lines mark the
gravitational mass of the individual neutron stars in those four
binary neutron star systems for which the individual masses have been
well established.\label{Fig3} }
\end{figure*}

\begin{table*}[b]
\caption{Mass data for double neutron star binaries}
\centerline{
\begin{tabular}{lcccccccc}
\hline \hline
Binary & Grav. Masses ($M_{\odot}$) & Combined Mass ($M_{\odot}$) &
A \tablenotemark{a} & D & L & UT & FPS & APR \\
\hline
J1518+4904 \tablenotemark{b} \tablenotemark{c} & 1.05, 1.56 & 2.62 $\pm$0.07  &  BH   &  HNS  & NS  & HNS & HNS & SNS \\
B1534+12 \tablenotemark{c}  & 1.339, 1.339 & 2.678 $\pm$0.012 &  BH   &  HNS  & NS  & HNS & HNS & SNS \\
B1913+16 \tablenotemark{c}  & 1.3874, 1.4411 & 2.8285 $\pm$0.0014 &  BH   &  BH   & NS  & HNS & BH  & HNS \\
B2127+11C \tablenotemark{c} & 1.349, 1.363 & 2.7122 $\pm$0.0006 &  BH   &  HNS  & NS  & HNS & HNS & SNS \\
B2303+46  \tablenotemark{b} \tablenotemark{c}  & 1.30, 1.34 & 2.64 $\pm$0.05 &  BH   &  HNS  & NS  & HNS & HNS & SNS \\
J0737-3039 \tablenotemark{d}& 1.250, 1.337 & 2.588 $\pm$0.003  &  HNS  &  HNS  & NS  & HNS & HNS & SNS \\
\hline
\end{tabular}}
\tablenotetext{a}{Possible outcomes of binary merger for each equation
of state according to computed total rest mass: (NS) neutron star; (SNS)
supramassive neutron star; (HNS) hypermassive neutron star; (BH) 
black hole.}
\tablenotetext{b}{For these systems the binary inclination angle has not
been determined precisely, so that the combined mass is known to much
higher accuracy than the individual neutron star masses.}
\tablenotetext{c}{Thorsett \& Chakrabarty (1999).  Lower and upper
limits for comined mass are $95\%$ central limits.}
\tablenotetext{d}{Lyne et.~al.~(2004).}
\label{Table2}
\end{table*}

For a polytropic EOS, on the other hand, $n$ is strictly independent
of the density.  We can therefore expect good agreement between
polytropic and nuclear models only as long as $n_{\rm eff}$ is close
to $n_{\rm nucl}$ for most of the matter in the star.  As can be seen
in the panel in Figure \ref{Fig2}, $n_{\rm eff}$ is very close to
$n_{\rm FPS}$ for densities in the range $(1.5 - 3) \times 10^{15}
\mbox{g} \mbox{cm}^{-3}$.  This coincides quite well with the regime
of maximum densities in equilibrium models for which the agreement
between the polytropic and nuclear models is fairly good.  For lower
densities, $n_{\rm FPS}$ drops significantly below $n_{\rm eff}$, and
it is therefore not surprising that stellar models with maximum
densities in this regime no longer agree very well.

For all other EOSs we find very similar results.  In all cases, the
nuclear adiabatic index $n_{\rm nucl}$ is very close to the effective
polytropic index $n_{\rm eff}$ (as determined from the central
condensation) for densities close the the maximum density of the
maximum mass TOV star.  Matter in this density regime dominates the
structure of the star for uniform and moderate degrees of differential
rotation, leading to good agreement between nuclear and polytropic
models.  For larger degrees of differential rotation, the star becomes
increasingly ``toroidal'' and bloated, and the maximum density shifts
to smaller values.  As a consequence, the star is dominated by matter
at smaller densities, for which $n_{\rm nucl}$ no longer agrees with
$n_{\rm eff}$.  Accordingly, the results for the polytropic models are
no longer in good agreement with the nuclear models.

%
%

\section{Summary and Discussion}
\label{summary}

We have constructed models of differentially rotating neutron stars in
general relativity for a sample of realistic nuclear equations of
state.  We find that their maximum rest masses are typically larger
than those of non-rotating neutron stars by about 50\%.  This increase
is significantly less than the increases for moderately stiff
polytropic equations of state (compare LBS).  This deviation can be
explained in terms of a drop in the maximum density for stars with
large degrees of differential rotation.  At these smaller densities,
the stiffness of the nuclear equation of state, as determined from
equation (\ref{pol_nucl}), is different from that at the maximum
density of maximum mass nonrotating or uniformly rotating stars.
Consequently, predictions based on simple polytropic models break down
for sufficiently large degrees of differential rotation.

Even a mass increase of 50\% may well be sufficient to support the
remnant of binary neutron star coalescence.  In fact, our findings can
be used to predict possible fates for the six known double neutron
star binaries as listed in Table \ref{Table2}.  For four of those,
B1524+12, B1913+16, B2127+11C and the newly discovered binary
J0737-3039, the individual gravitational masses are known to high
accuracy (Thorsett \& Chakrabarty 1999; Burgay et.~al.~2003, Lyne
et.~al.~2004).  For the two remaining ones, J1518+4904 and B2303+46,
the orbital inclination is unknown, so that the combined gravitational
mass is known to much higher accuracy than the individual masses.
Assuming no mass loss during the coalescence and merger (Faber,
Grandcl\'ement \& Rasio 2003) the total rest mass of the binary is
conserved (the total gravitational mass is not conserved, due to
neutrino and gravitational radiation losses).  The rest mass
corresponding to a particular gravitational mass depends on the
equation of state, as is shown in Figure \ref{Fig3} for the four
binaries with well-established individual masses.

The total rest mass $M_0^{\rm tot}$ of the binary is an indicator of
the fate of the merger remnant.  If the total rest mass is less than
the maximum (nonrotating) TOV mass $M_0^{\rm TOV}$, then the remnant
can be supported against gravitational collapse even without rotation.
This possibility is identified by ``NS'' in Table \ref{Table2}.  If
$M_0^{\rm tot}$ is greater than $M_0^{\rm TOV}$, but less than the
maximum mass of uniformly rotating stars $M_0^{UNI}$, then the merger
could lead to a supramassive star, supported by uniform rotation
(denoted by ``SNS'' in Table \ref{Table2}).  If $M_0^{\rm tot}$ is
greater than $M_0^{\rm UNI}$ but less than the maximum mass we find
for differentially rotating stars $M_0^{\rm DIF}$, merger may lead to
a hypermassive neutron star supported by differential rotation
(denoted by ``HNS'').  If, finally, $M_0^{\rm tot}$ is greater than
$M_0^{\rm DIF}$, the remnant will promply collapse to a black hole
(``BH'').  Clearly, these limits are only approximations, and the true
outcome will depend not only on the mass and EOS but also on the
non-equilibrium dynamics of the merger.

Since the rest masses of the observed binaries depend on the equation
of state, so do the possible fates.  Moreover, since all binary
neutron star systems known to date have similar total masses, their
fate, for a given equation of state, is presumably similar.  In
Table \ref{Table2}, the predictions for four of the six binaries are
identical; only the high mass binary B1913+16 and the low mass binary
J0737-3039 show any differences for some of the equations of state.

Table \ref{Table2} suggests two immediate conclusions.  For all the
recent equations of state we survey and for all binaries except for
B1913+16 the coalescence is likely to lead to a neutron star.  For
three out of the six equations of state, the merger of these binaries
could lead to a hypermassive neutron star.  Unlike supramassive
neutron stars, hypermassive neutron stars last only a fairly short
time, until magnetic fields or viscous dissipation drive the objects
toward uniform rotation, triggering a delayed collapse to a black hole
(see BSS).

Secondly, Table \ref{Table2} shows that the outcome of binary neutron
star coalescence is quite sensitive to the equation of state.  This
means that {\em the observational identification of a merger remnant,
and its possible delayed collapse to a black hole, could place a tight
constraint on the equation of state}.  Advanced LIGO (the Laser
Interferometer Gravitational Wave Observatory) and other new
gravitational wave laser interferometers may be able to observe binary
neutron star coalescence.  Comparing the most recent equations of
state, UT, FPS and APR, it seems likely that the coalescence will lead
to a rotationally supported neutron star.  Observation of the
remnant's delayed collapse to a black hole, or its absence, would then
help to put contraints on possible equations of state.

\acknowledgments

It is a pleasure to thank G.~Ravenhall and V.~J.~Pandharipande for
providing the data for equation of state APR.  This paper was
supported in part by NSF Grant PHY-0139907 at Bowdoin College and NSF
Grants PHY-0090310 and PHY-0205155 and NASA Grant NAG 5-10781 at the
University of Illinois at Urbana-Champaign.

%
%

\begin{appendix}

\section{Numerical Results for Maximum Masses}
\label{tables}

We list in the Tables below values for the maximum rest masses for
uniformly and differentially rotating polytropes.  For each equation
of state we tabulate the differential rotation parameter $\hat
A^{-1}$, the maximum energy density $\epsilon_{\rm max}$, the ratio of
the central and equatorial angular velocity $\Omega_c/\Omega_e$, the
ratio between polar and equatorial radii $R_p/R_e$, the ratio of the
(relativistic) rotational kinetic energy and the gravitational binding
energy $T/|W|$, the maximum rest mass $M_0^{\rm max}$, and the
fractional rest mass increase $(\delta M_0/M_0)$.  We also construct
differentially rotating polytropes for effective polytropic indices
$n_{\rm eff}$ as described in Section \ref{results} and listed in each
Table heading.  We include in the Tables the fractional mass increases
$(\delta M_0/M_0)_{\rm poly}$ for these polytropes.

All models are computed with 64 zones both in the radial and angular
direction, and with the Legendre polynomial expansion truncated at
$\ell = 16$ (see CST1 for details of the numerical implementation).
The error in the ratio $R_p/R_e$ is determined by our stepsize of
0.015 in this ratio (compare Section \ref{em}).  We finally note that
highly toroidal configurations depend very sensitively on the input
parameters, so that those mass increases depend on the parameters
adopted in our restricted survey.


\begin{table}
\caption{A: $M_0^{\rm TOV} = 1.92M_{\odot}$, $\overline{\epsilon} /
\epsilon_c = 0.326$, $n_{\rm eff}=0.621$}
\centerline{
\begin{tabular}{cccccccc}
\hline \hline
$\hat A^{-1}$  &
$\epsilon_{\rm max}/10^{15}$ g cm$^{-3}$ &
$\Omega_c / \Omega_e$  &
$R_p / R_e$  &
$T / \vert W \vert$  &
$M_0^{\rm max} / M_{\odot}$ &
$(\delta M_0 / M_0)$  &
$(\delta M_0 / M_0)_{\rm poly}$  \\
\hline
0.0 & 3.55 &  1.000 &  0.565 & 0.120 & 2.24 & 0.17 & 0.19 \\
0.3 & 3.45 &  1.465 &  0.460 & 0.180 & 2.50 & 0.30 & 0.34 \\
0.5 & 1.87 &  2.026 &  0.415 & 0.227 & 2.69 & 0.40 & 0.56 \\
0.7 & 0.938 & 2.448 &  0.220 & 0.293 & 2.90 & 0.51 & 0.76 \\
1.0 & 1.12 &  3.584 &  0.310 & 0.235 & 2.34 & 0.22 & 0.43 \\
\hline
\end{tabular}}
\end{table}


\begin{table}
\caption{D: $M_0^{\rm TOV} = 1.89M_{\odot}$, $\overline{\epsilon} /
 \epsilon_c = 0.284$, $n_{\rm eff} = 0.738$}
\centerline{
\begin{tabular}{cccccccc}
\hline \hline
$\hat A^{-1}$  &
$\epsilon_{\rm max}/10^{15}$ g cm$^{-3}$   &
$\Omega_c / \Omega_e$  &
$R_p / R_e$  &
$T / \vert W \vert$  &
$M_0^{\rm max} / M_{\odot}$ &
$(\delta M_0 / M_0)$  &
$(\delta M_0 / M_0)_{\rm poly}$  \\
\hline
0.0 & 2.73 &  1.000 & 0.565 & 0.110     & 2.22 & 0.18 & 0.18 \\
0.3 & 2.64 &  1.344 & 0.505 & 0.153     & 2.41 & 0.28 & 0.28 \\
0.5 & 1.51 &  1.963 & 0.385 & 0.235     & 2.95 & 0.56 & 0.55 \\
0.7 & 0.884 & 2.455 & 0.275 & 0.279     & 3.02 & 0.60 & 1.0  \\
1.0 & 0.774 & 3.463 & 2.965 & 0.224E-07 & 2.97 & 0.57 & 0.66 \\
\hline
\end{tabular}}

\end{table}


\begin{table}
\caption{L: $M_0^{\rm TOV} = 3.23M_{\odot}$, $\overline{\epsilon} /
 \epsilon_c = 0.339$, $n_{\rm eff} = 0.586$}
\centerline{
\begin{tabular}{cccccccc}
\hline \hline
$\hat A^{-1}$  &
$\epsilon_{\rm max}/10^{15}$ g cm$^{-3}$   &
$\Omega_c / \Omega_e$  &
$R_p / R_e$  &
$T / \vert W \vert$  &
$M_0^{\rm max} / M_{\odot}$ &
$(\delta M_0 / M_0)$  &
$(\delta M_0 / M_0)_{\rm poly}$ \\
\hline
0.0 & 1.26  & 1.000 & 0.550 & 0.134 & 3.87 & 0.20 & 0.19 \\
0.3 & 1.13  & 1.505 & 0.425 & 0.210 & 4.52 & 0.40 & 0.37 \\
0.5 & 0.498 & 1.970 & 0.320 & 0.273 & 4.78 & 0.48 & 0.52 \\
0.7 & 0.378 & 2.502 & 0.010 & 0.298 & 5.04 & 0.56 & 0.69 \\
1.0 & 0.374 & 3.464 & 0.025 & 0.272 & 4.06 & 0.26 & 0.35 \\
\hline
\end{tabular}}

\end{table}


\begin{table}
\caption{UT: $M_0^{\rm TOV} = 2.17M_{\odot}$, $\overline{\epsilon} /
 \epsilon_c = 0.321$, $n_{\rm eff}=0.635$}
\centerline{
\begin{tabular}{cccccccc}
\hline \hline
$\hat A^{-1}$  &
$\epsilon_{\rm max}/10^{15}$ g cm$^{-3}$   &
$\Omega_c / \Omega_e$  &
$R_p / R_e$  &
$T / \vert W \vert$  &
$M_0^{\rm max} / M_{\odot}$ &
$(\delta M_0 / M_0)$  &
$(\delta M_0 / M_0)_{\rm poly}$ \\
\hline
0.0 & 2.73  & 1.000 & 0.565  & 0.122 & 2.56 & 0.18 & 0.19 \\
0.3 & 2.66  & 1.455 & 0.460  & 0.182 & 2.86 & 0.32 & 0.33 \\
0.5 & 2.02  & 2.145 & 0.490  & 0.195 & 2.91 & 0.34 & 0.57 \\
0.7 & 0.763 & 2.441 & 0.220  & 0.293 & 3.25 & 0.50 & 0.81 \\
1.0 & 0.744 & 3.468 & 0.010 & 0.273  & 2.97 & 0.37 & 0.47 \\
\hline
\end{tabular}}

\end{table}


\begin{table}
\caption{FPS: $M_0^{\rm TOV} = 2.10M_{\odot}$, $\overline{\epsilon} / 
\epsilon_c = 0.316$, $n_{\rm eff}=0.648$}
\centerline{
\begin{tabular}{cccccccc}
\hline \hline
$\hat A^{-1}$  &
$\epsilon_{\rm max}/10^{15}$ g cm$^{-3}$   &
$\Omega_c / \Omega_e$  &
$R_p / R_e$  &
$T / \vert W \vert$  &
$M_0^{\rm max} / M_{\odot}$ &
$(\delta M_0 / M_0)$  &
$(\delta M_0 / M_0)_{\rm poly}$ \\
\hline
0.0 & 2.92 & 1.000 & 0.565 & 0.117 & 2.45 & 0.17 & 0.19 \\
0.3 & 2.83 & 1.440 & 0.475 & 0.172 & 2.72 & 0.29 & 0.32 \\
0.5 & 1.36 & 1.942 & 0.360 & 0.248 & 3.08 & 0.46 & 0.59 \\
0.7 & 0.836 & 2.446 & 0.275 & 0.278 & 3.06 & 0.45 & 0.82 \\
1.0 & 0.748 & 3.469 & 0.010 & 0.275 & 2.99 & 0.42 & 0.47 \\
\hline
\end{tabular}}

\end{table}


\begin{table}
\caption{APR: $M_0^{\rm TOV} = 2.67 M_{\odot}$, $\overline{\epsilon} /
 \epsilon_c = 0.377$, $n_{\rm eff} = 0.495$}
\centerline{
\begin{tabular}{cccccccc}
\hline \hline
$\hat A^{-1}$  &
$\epsilon_{\rm max}/10^{15}$ g cm$^{-3}$   &
$\Omega_c / \Omega_e$  &
$R_p / R_e$  &
$T / \vert W \vert$  &
$M_0^{\rm max} / M_{\odot}$ &
$(\delta M_0 / M_0)$  &
$(\delta M_0 / M_0)_{\rm poly}$ \\
\hline
0.0 & 2.44 &  1.000 & 0.580  & 0.132 & 3.10 & 0.16 & 0.20 \\
0.3 & 1.77 &  1.553 & 0.445  & 0.210 & 3.49 & 0.31 & 0.42 \\
0.5 & 1.06 &  1.970 & 0.370  & 0.245 & 3.34 & 0.25 & 0.76 \\
0.7 & 0.770 & 2.449 & 0.280  & 0.275 & 3.19 & 0.19 & 0.46 \\
1.0 & 0.691 & 3.451 & 0.010  & 0.276 & 3.12 & 0.17 & 0.81 \\
\hline
\end{tabular}}

\end{table}

\end{appendix}

\end{document}